\documentclass{llncs}

\usepackage{hyperref}

\begin{document}

\title{Multi-Agent Programming Contest 2011 \\ --- \\ The Python-DTU Team}

\author{J{\o}rgen Villadsen \and Mikko Berggren Ettienne \and Steen Vester}

\institute{Department of Informatics and Mathematical Modelling \\
Technical University of Denmark \\
Richard Petersens Plads, Building 321, DK-2800 Kongens Lyngby, Denmark}

\maketitle
\thispagestyle{plain} \pagestyle{plain}

\smallskip

\begin{abstract}
We provide a brief description of the Python-DTU system, including the overall design, the tools and the algorithms that we plan to use in the agent contest.

\

Updated 1 October 2011: Appendix with comments on the contest added.
\end{abstract}

\smallskip

\section*{Introduction}

\begin{enumerate}
\item
The name of our team is Python-DTU. We participated in the contest in 2009 and 2010 as the Jason-DTU team \cite{Boss+2010,Vester+2011},
where we used the Jason platform \cite{Bordini+2007}, but this year we use just the programming language Python.
\par\medskip
We intend in a later paper to elaborate on the reasons for abandoning the Jason platform and its agent-oriented programming language Agent\-Speak.
\medskip
\item
The members of the team are as follows:
\begin{itemize}
\smallskip
\item J{\o}rgen Villadsen, PhD
\smallskip
\item Mikko Berggren Ettienne, MSc student (new in the team this year)
\smallskip
\item Steen Vester, MSc student
\smallskip
\end{itemize}
We are affiliated with DTU Informatics (short for Department of Informatics and Mathematical Modelling, Technical University of Denmark, and located in the greater Copenhagen area).
\medskip
\item
The main contact is associate professor J{\o}rgen Villadsen, DTU Informatics, email: \email{jv@imm.dtu.dk}
\medskip
\item
We expect that we will have invested approximately 400 man hours when the tournament starts.
\end{enumerate}

\section*{System Analysis and Design}

\begin{enumerate}
\item
The competition is built on the Java MASSim platform and the Java EISMASSim framework is distributed with the competition files.
This framework is based on EIS and abstracts the communication between the server and the agents to simple Java method calls and callbacks.
\par\medskip
To utilize this framework we started out with the Java implementation of Python called Jython which in contrast to Python can import Java libraries and classes.
To support agent communication in our multi-agent system we have so far used the Apache ActiveMQ as a messaging server which offers clients for all popular programming languages. 
Using the EISMASSim Java framework together with ActiveMQ clients written in Python gluing it all together with Jython gave some performance issues when exchanging percepts between the agents.
\par\medskip
We found that each component performed well tested in a controlled context which suggested that the issues were related to the interaction between the components.
\par\medskip
We decided to skip Jython and EISMASSim to instead follow a much cleaner Python-only implementation.
Even though some work was needed to implement the protocol specific parts which EISMASSim handled, this left us with a more flexible implementation
of which we have complete knowledge and control of every part of the implementation.
It also solved the performance issues related to component interaction.
\par\medskip
We also plan to skip the messaging server and instead let the agents communicate directly using a simple and efficient text-based protocol to further improve the performance of the system.
A simple text-based protocol allows us to keep the distributivity and modularity.
Thus it will still be posible to use agents written in different programming languages.
Furthermore, by implementing our own message server, we can tweak the low-level features to suit the need of our specific system.
\medskip
\item
We do not use any existing multi-agent system methodology.
\medskip
\item
We do not plan to distribute the agents on several machines.
\medskip
\item
We do not plan a solution with a centralization of coordination/information on a specific agent.
Rather we plan a decentralized solution where agents share percepts through messages and coordinate actions using distributed algorithms.
\medskip
\item
The team communication is based on the publisher-subscriber pattern.
Our message server has a number of topics to which agents can subscribe and publish messages.
This supports one-to-one communication and one-to-many communication in a simple way where the agents determine which topics to subscribe to.
\medskip
\item
To assign goals to agents we use a ring-based auction algorithm.
Each round in an auction includes $n$ messages where $n$ is the number of agents participating.
Before participating in an auction all agents score their top $n$ goals and use this score to determine their bidding strategy.
The algorithm terminates when all $n$ agents are assigned to a unique goal. 
\medskip
\item
Each agent acts on its own behalf based on its local view of the world which is updated through percepts and is thus autonomous and reactive.
This is implemented as an agent-control-loop in which the agents decide which actions to execute based on their current view of the world.
We have considered implementing an algorithm to determine the best way to parry a series of attacks from a saboteur agent which would make the agents proactive.
\end{enumerate}

\section*{Software Architecture}

\begin{enumerate}
\item
We implement the multi-agent system using just the programming language Python.
\par\medskip
Even though we all have more experience with Java, we choose Python as our programming language, as we think it has some advantages over Java, mainly in development speed/programmer effectiveness.
Some of the reasons being that Python in contrast to Java:
\begin{itemize}
\smallskip
\item is dynamically typed
\smallskip
\item is concise
\smallskip
\item is compact
\smallskip
\item supports multiple programming paradigms (object-oriented, imperative, functional)
\smallskip
\item is popular for scripting
\smallskip
\item does not need to be compiled before execution
\smallskip
\end{itemize}
\medskip
\item
We use Python 2.7 on Ubuntu Linux and Mac OS X as the development platforms and GEdit, Eclipse and TextMate as code editors/IDEs.
\medskip
\item
As the runtime platform for the competition we plan to use a suitable Linux system with Python 2.7.
\medskip
\item
We plan to use the following algorithms:
\begin{itemize}
\smallskip
\item All-pair shortest path extended to support dynamic vertex addition
\smallskip
\item Custom breadth-first-searches
\smallskip
\item Ring-based agreement algorithm
\smallskip
\end{itemize}
\end{enumerate}

\

\section*{Acknowledgements}

Thanks to Niklas Christoffer Petersen, MSc student, for comments.

\vfill

\begin{center}
More information about the Python-DTU team is available here:
\\[3ex]
\url{http://www.imm.dtu.dk/~jv/MAS}
\end{center}

\

\section*{Appendix}

Our idea to skip the message server and implement our own text-based protocol was slightly modified during the implementation phase.
Instead of implementing our own message system, we reached the pragmatic conclusion to instead let the agents communicate trough direct reference to a shared data structure.
This allowed us to spend more time on other important issues and freed us from all performance issues.
The main reason for the message system idea was to preserve modularity and this could easily be implemented at a later state if time allowed. 

In the early development phase we suspected our dynamic all-pair shortest path algorithm as a candidate for our performance issues. 
However we discovered that the issues were I/O related, and found that we had plenty of processing time and power to perform multiple stock graph search algorithms at each simulation step.
This made our dynamic all-pair shortest path algorithm superfluous for this scenario.
However it could be highly relevant in case of less processing time or bigger graphs.

\end{document}